\documentclass[]{spie}  
\usepackage[]{graphicx}

\title{Beyond the standard approximations:  an analysis leading to a correct description of phase instabilities in semiconductor lasers} 
\author{Gil L.\supit{a,b} and Lippi G.L.\supit{a,b}
\skiplinehalf
\supit{a}Institut Non Lin\'eaire de Nice, Universit\'e de Nice-Sophia Antipolis, Valbonne, France \\
\supit{b}CNRS, UMR 7335, 1361 Route des Lucioles, F-06560 Valbonne, France
}

\authorinfo{Correspondence to L. Gil\\E-mail: lionel.gil@inln.cnrs.fr}
 
\begin{document} 
\maketitle 

\begin{abstract}
Following an overview of modeling of (longitudinal) multimode semiconductor laser dynamics, we analyze in detail a model proposed in 2006 to explain deterministic, phase-locked modal alternation, experimentally observed a decade ago.  Through a stability analysis, we prove that the numerically obtained electromagnetic field evolution, interpreted as an explanation of the experiments, is nothing more than an extremely long transient, so long as to be hardly identifiable in an entirely numerical approach.  Comparison with a model we have recently derived, which predicts a phase instability (Benjamin-Feir-like) compatible with the experimental observations, highlights the crucial ingredient for the dynamics.  The wide spectrum of unstable eigenvalues accompanying the phase instability plays the role of an equivalent noise in a fully deterministic description, thus reconciling the heuristic models which could qualitatively reproduce the experimental observation either with deterministic equations in the presence of mode-coupling, or through stochastically driven modal decompositions.  
\end{abstract}

\keywords{Semiconductor lasers, Phase Instability, laser modeling, Complex Ginzburg-Landau Equation, Laser parameters, Benjamin-Feir Instability, Codimension 2 bifurcation, Noise sources}

\section{INTRODUCTION}
\label{sec:intro}  

The dynamics of multimode lasers has occupied a prime place in the investigation of laser behaviour since, starting from their first realizations~\cite{Maiman1960}, lasers operated on several longitudinal (and at times transverse) cavity modes.  Indeed, only with further work did it become possible to restrict the operation to one longitudinal and transverse mode oscillation.  Thus, questions on the correct modeling of the interaction between several (longitudinal) modes and the active medium were vigorously debated very early on.  One of the first and most successful proposals for a satisfactory description of laser dynamics came from Tang, Statz and De Mars~\cite{Tang1963} who introduced a model -- widely used up until today -- which treats the individual cavity mode intensities as directly coupled to the energy reservoir (population inversion), independently of one another.  The question of mutual coupling among modes was effectively solved in a later seminal paper~\cite{Fleck1964}.

Since the early work, multimode dynamics has generally been extensively studied either because of its intrinsic features, for the regular or irregular temporal oscillations which accompany it, or because of the disruption that it may cause whenever high spectral purity is required.  One particular regime has, however, attracted particular attention because of its peculiar features:  the so-called antiphase regime.  Under this mode of operation, several longitudinal laser modes oscillate, but are mutually coupled in such a way as to phase-lock, thereby maintaining a constant total output power.
In the simplest configuration such a behaviour was observed in two-mode lasers~\cite{Lenstra1980,Zeghlache1988}, but its full impact was appreciated only with the very first attempts at building efficient frequency-doubled lasers~\cite{Baer1986}, followed by detailed investigations of the dynamical properties of antiphase states~\cite{Wiesenfeld1990}.

The modal picture is somewhat different when one examines the operation of semiconductor lasers.  Devices which are capable of multi-longitudinal mode operation (i.e., in-plane geometries, whether bulk, guided, or Quantum Well junctions) are extremely sensitive to any form of external feedback.  Aside from large levels of re-injection, introduced for stabilization purposes, practically any amount of reinjection -- no matter how small -- causes a destabilization of the laser output which manifests itself in the form of Low-Frequency-Fluctuations~\cite{Huyet1998}, Mode-Hopping~\cite{Gray1991}, Coherent Collapse~\cite{Lenstra1985}, and fast phase jumps~\cite{Vashenko1998} with antiphase oscillations during the fast transient which wash out the details of the dynamics~\cite{Huyet1999}.  An early overview~\cite{Tkach1986} of the different regimes, which predates their detailed investigations, and a summary~\cite{Wallace2000} provide a useful global picture.  

Noise dominates the multimode semiconductor laser dynamics and, as such, models have concentrated on the most appropriate ways of introducing its influence on the dynamical evolution. A thorough analysis of the influence of noise on the operation of a semiconductor laser and on the associated fluctuations was reported by Henry~\cite{Henry1986}, while a review of the different physical sources of noise in a semiconductor laser has been given by Osi\'nski and Buus~\cite{Osinski1987}.
Phenomenological multimode rate equations have been devised~\cite{Copeland1983,Henry1984,Wentworth1990} which account for mode partition~\cite{Linke1985,Ohtsu1989} (a regime where the instantaneous laser power is randomly distributed among several simultaneous modes) or mode hopping~\cite{Ohtsu1989,Ohtsu1986} (instantaneous oscillation on an individual mode, with random jumps among modes).

In 2004 new experimental results~\cite{Yacomotti2004} demonstrated that the general picture, put together over the course of two decades, where multimode semiconductor laser dynamics is driven and controlled by noise, was incomplete.  Indeed, measurements performed on a Multiple Quantum Well (MQW) device conclusively showed that a new kind of dynamics exists, at least for some devices, where the laser emission is composed of a regular, cyclic alternation of modes phase-locked in such a way as to maintain constant the power output by the device~\cite{Yacomotti2004,Furfaro2004}.  Among various other oscillatory regimes with mixed features (e.g., partial modulation of the total laser power), the most unusual one corresponds to a modal switching dynamics nearly devoid of a modulation of the total output, but with full modulation in the modal, frequency-resolved emission, with a sequence starting from the bluest mode and evolving towards the reddest mode; once the last mode in the active range is reached, the sequence restarts from the bluest mode following the same regular features (with emission times and mode transitions which depend on the pairs of modes)~\cite{Yacomotti2004,Furfaro2004}.   Another experiment followed~\cite{Tanguy2006} which showed a dynamical regime with substantially the same features, although much more complex, in an entirely different device:  a multilongitudinal mode Quantum Dot Laser (QDL).  These observations proved that the deterministic regime is not exclusive to MQW devices and that its origin must be sought outside the physical characteristics of MQWs.  

The regular, deterministic dynamics entirely contrasted with the stochastic picture of all previous observations and demanded a new kind of physical description for its interpretation.  To this end, new phenomenological multimode rate equation models were constructed, which included the interaction among modes through Four-Wave-Mixing (FWM)~\cite{Yacomotti2004}, Cross-Saturation (CS) and Self-Saturation (SS)~\cite{Ahmed2003}.  In this approach, coupling among modes is phenomenologically introduced on the basis of physical considerations, and a satisfactory reproduction of the main observations is achieved without the need for introducing noise.  Thus, determinism is recovered through the introduction of mode-coupling and excluding noise from the mathematical description of the system.

An interesting alternative approach~\cite{Ahmed2002}, cited also in the first experimental report~\cite{Yacomotti2004}, makes a choice opposite to the usual one~\cite{Copeland1983,Henry1984,Wentworth1990} by introducing noise at the level of the global electric field (i.e., the total field contributed by all modes).  This choice automatically introduces a form of intermodal coupling, through noise, which is thereby partitioned onto each modal intensity.  In this approach, while the {\it global} noise is a random process, the {\it individual noise component}, i.e. the instantaeous noise projected onto each individual mode, introduces a correlation between the fluctuations of different modes.  This choice is unusual and shows that determinism can be recovered through correlations, even though the underlying source of correlation is a random process.

Summarizing, some kind of laser devices are capable of showing deterministic modal dynamics, accompanied, in some regimes, by a constant power output, which indicates modal phase-locking.  Modeling has been achieved either through multimode deterministic rate equations modified to phenomenologically include mode coupling, or though equivalent multimode equations (without mode coupling) including noise introduced for the global field (partitioned among the individual modes).  

Following the publication of the experimental results~\cite{Yacomotti2004,Furfaro2004}, a more advanced form of modelling was introduced where the medium's susceptibility, which can be expressed in an approximate analytical form~\cite{Balle1998}, was explicitely used to obtain coupled Partial Differential Equations (PDEs) which describe the evolution of the electric field and of the carrier density~\cite{Serrat2006}.  This modeling choice represents the most sophisticated approach developed up until our contribution~\cite{Gillong} and contains several similarities.  We are therefore going to closely analyse this model, interpret its different components, comment on its predictions and compare them to those we obtain~\cite{Gil2014}.

\section{The Serrat--Masoller (SM) model}

The model derived by Serrat and Masoller~\cite{Serrat2006} describes the evolution of the laser field, in interaction with the carriers of a semiconducting material, in a bidirectional ring cavity.  Their starting point is the same as the one we take~\cite{Gil2014,Gillong}, i.e., the wave equation for the electromagnetic (e.m.) field coupled to a formally described polarization for the source term, together with the evolution equation for the carrier density.  However, the Adiabatic Elimination (AE) of the polarization variable and the Slowly Varying Envelope Approximation (SVEA), together with the Rotating Wave Approximation (RWA), are assumed to hold~\cite{Serrat2006}.  Thus, the equations for the slowly varying  field amplitudes, in the two counterpropagating directions, are directly written out.  In addition, the frequency dependence of the susceptibility is described phenomenologically in the parabolic approximation~\cite{Serrat2006} thereby fixing all the physical functional dependencies.  The parabolic approximation, commonly used for semiconductor models~\cite{Coldren1995}, forces a symmetric lineshape around the lasing frequency and thereby misses a crucial point (cf. eq.~(\ref{betadef})) in the description of the phase instability (cf. eqs.~(\ref{eigenv}-\ref{alphabeta})).  Periodic boundary conditions are introduced~\cite{Serrat2006}, together with finite mirror reflectivity, to take into account the role of the (ring) cavity.

Rewritten in a notation more similar to ours, the model~\cite{Serrat2006} reads:
\begin{eqnarray}
\label{SerratEquations1}
{{\partial F}\over{\partial t}} & = & -{{\partial F}\over{\partial z}}+ \mathcal{G} \left[ -i \alpha N+\left(N-1\right)\left(1+G_{d}{{\partial^2}\over{\partial z^2}}\right)-\gamma_{int}\right]F \, , \\
\label{SerratEquations2}
{{\partial B}\over{\partial t}} & = & +{{\partial B}\over{\partial z}}+ \mathcal{G} \left[ -i \alpha N+\left(N-1\right)\left(1+G_{d}{{\partial^2}\over{\partial z^2}}\right)-\gamma_{int}\right]B \, , \\
\label{SerratEquations3}
{{\partial N}\over{\partial t}} & = & j-\gamma_{n}N-\left(N-1\right)\left(\vert F \vert^2+\vert B \vert^2\right)+d{{\partial^2N}\over{\partial z^2}} \, , 
\end{eqnarray}
where $F$ and $B$ represent the slowly-varying amplitudes of the forward and backward propagating e.m. field, $\mathcal{G}$ stands for normalized linear gain, $\alpha$ is the linewidth enhancement factor, $G_d$ is the contribution of the susceptibility in the parabolic approximation, $\gamma_{int}$ represents the non-radiative losses for the field, $j$ stands for the injected current density (in normalized units) and $d$ plays the role of a normalized diffusion constant.

It is important to notice a fundamental difference between the results that we show below and those previously reported~\cite{Serrat2006}, where the authors performed a reduction of the predicted field evolution by projecting the total intensity onto a set of cavity modes.  As specifically stated in the original paper~\cite{Serrat2006} ``In the multimode regime the output power exhibits fast oscillations due to mode beating" removed by convolving the predicted (total) field evolution with a reference field centered at each modal frequency.  It is the sum of the resulting set of modal intensities -- labelled~\cite{Serrat2006} {\it incoherent sum} -- which was shown to display a behaviour similar to that of the experiments~\cite{Yacomotti2004,Furfaro2004} (mode alternation with ``phase locking").  This choice of information reduction was justified~\cite{Serrat2006} by the limited bandwidth encountered in experiments (at least at the time when they were conducted) which prevented the detection of the intermode beats.  

\section{Predictions from the S-M model}\label{SMmodel}

The system of equations (\ref{SerratEquations1}-\ref{SerratEquations3}) possesses a 4-continuous-parameters ($\phi_{F}$, $\phi_{B}$, $\theta$ and $K$) family of stationary solutions
\begin{eqnarray}
\label{pos1}
F & = & R \, cos(\theta) \, e^{i\left(\Omega t-K z+\phi_{F}\right)} \, , \\
\label{pos2}
B & = & R \, sin(\theta) \, e^{i\left(\Omega t+K z+\phi_{B}\right)} \, , \\
\label{pos3}
N & = & N_{0} \, ,
\end{eqnarray}
where $\phi_{F}$ and $\phi_{B}$ represent the phases of the solutions, $\theta$ the phase angle of the vector $R e^{i \theta}$ which represents the amplitude of the e.m. field (decomposed into the forward, $F$, and backwards, $B$, directions), and $K$ the wavevector mismatch, with the following values obtained from the formal solution:
\begin{eqnarray}
\label{solutionSerrat1}
R^2 & = & {{G_{d}\left(\gamma_{n}-j\right) K^2-\gamma_{n}\left(\gamma_{int}+1\right)+j}\over{\gamma_{int}}} \, , \\
\label{solutionSerrat2}
N_{0} & = & {{G_{d}K^2-\gamma_{int}-1}\over{G_{d}K^2-1}} \, , \\
\label{solutionSerrat3}
\Omega & = & {{G_{d}K^3-G_{d}\alpha {\mathcal{G}} K^2-K+\alpha {\mathcal{G}} \left( 1+\gamma_{int} \right) }\over{G_{d}K^2-1}} \, , 
\end{eqnarray}
$N_0$ being the carrier stationary amplitude and $\Omega$ the angular frequency mismatch relative to that of the (rotating) reference frame.

We are now interested in the linear stability of these solutions. For the sake of clarity, we restrict ourselves to the case
where $0$$=$$\phi_{F}$$=$$\phi_{B}$$=$$K$, but situations with $K$$\ne$$0$ are not fundamentally different.

Among the linear eigenvalues, three are critical and take the form
\begin{equation}
\begin{array}{llrr}
\lambda_{1}&=&-i k &-r_{12} k^2+ \ldots \quad ,
\cr
\lambda_{2}&=&+i k &-r_{12} k^2+ \ldots \quad ,
\cr
\lambda_{3}&=&i \left(1-2 cos^2(\theta) \right) k&-r_{3}k^2+ \ldots \quad , 
\end{array}
\end{equation}
where $k$ is the perturbation wavenumber and the dots stand for higher-order powers of $k$. $r_{12}$ and $r_{3}$ are real numbers defined as
\begin{eqnarray}
\label{coeff12}
r_{12} & = & {\mathcal{G}} \gamma_{int} G_{d} \, , \\
\label{coeff3}
r_{3} & = & {{2\left(\gamma_{n}-j\right)\left(cos^2(\theta)-cos^4(\theta)\right)+G_{d}\gamma_{int}^3 {\mathcal{G}}^2\left(
\gamma_{int}\gamma_{n}+\gamma_{n}-j\right)}\over{\gamma_{int} {\mathcal{G}} \left(\gamma_{int}\gamma_{n}+\gamma_{n}-j\right)}} \, .
\end{eqnarray}
For the parameter values used in the simulations~\cite{Serrat2006} these coefficients are positive, thus the solutions (eqs.~(\ref{pos1}-\ref{solutionSerrat3})) are stable with respect to the phase.  In a numerical box of length $L$, the phase gradients are then expected to decrease with a characteristic (normalized) time $\tau$ given by
\begin{equation}
\tau={{1}\over{r_{12} \left({{2\pi}\over{L}}\right)^2}} \simeq 10^{5} \, ,
\end{equation}
expressed in the units of the original paper~\cite{Serrat2006}.

Without the support of the previous theoretical analysis (as was the case~\cite{Serrat2006}), this long characteristic time can lead to confusing the transient evolution with an asymptotic state. With the integration time step used in the original paper~\cite{Serrat2006}, $\triangle t=1/300$, the characteristic time $\tau$ is reached only after $\simeq$$3 \times 10^{7}$ iterations!  Hence, it is not surprising that -- especially with the standard computing means of the mid--2000's -- the very long transient may have been assumed to be a steady state and may have been interpreted as a phase instability.

\begin{figure}
\begin{center}
\begin{tabular}{c}
\includegraphics[height=7cm]{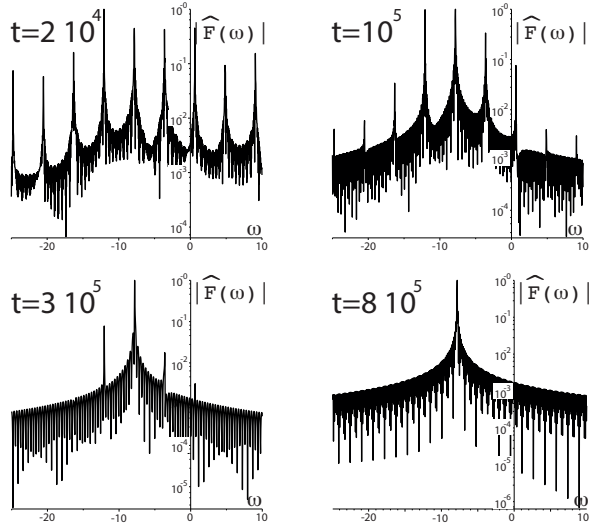}
\end{tabular}
\end{center}
\caption[fig03] 
{ \label{fig03} 
Numerical simulation of eqs.~(\ref{SerratEquations1}-\ref{SerratEquations3}) with exactly the same parameter values used for Fig 5.c~\cite{Serrat2006}. The plots show the power spectra of $F$ at different times: $2 \times 10^{4}$, $1 \times 10^{5}$, $3 \times 10^{5}$ and $8 \times 10^5$. The power spectra are computed by integration over a $10^{3}$ time interval (in rescaled units).  The spectrum gradually evolves with time, to finally become rigorously stationary for the last panel.}
\end{figure}

Fig.\ref{fig03} displays the long time evolution of the power spectrum of $F$ in the parameter regime for which the dynamics was interpreted~\cite{Serrat2006} as mode-switching.  We clearly see that the multimode dynamics gradually disappears to give rise to a monochromatic wave.  Thus, there is no possibility for the occurrence of a phase instability in this regime; however, we remark that the true monochromatic operation is reached only at extremely long times ($\tau = 8 \times 10^5$ in normalized units, i.e. $\simeq 2.4 \times 10^8$ integration time steps).

Note that the Complex Ginzburg-Landau equation (CGLE) is known to display Spatio-Temporal Intermittency (STI) even in the phase-stable regime~\cite{Chate1994}.  Indeed, if the initial conditions are sufficiently smooth, the CGLE converges to a regular continuous wave solution. On the contrary, if the initial conditions contain strong amplitude gradients, these will persist and produce an amplitude-turbulent regime.  Aware of this possibility, we have used strongly perturbed initial conditions for the numerical simulations as in Fig.~\ref{fig03} in order to check for the possible existence of STI in the SM model.  Even in this case, we obtain (a slow) convergence towards a stationary, monochromatic solution, thus excluding the possibility of an amplitude-turbulent regime, which would, in any case, be inconsistent with the forecasted constant {\it incoherent sum}~\cite{Serrat2006}.

\section{Codimension 2 (CD2) model}  

Let us outline the main features of the new derivation of a model based on a PDE description of the semiconductor laser.  
We consider a unidirectional (ring) laser by writing the propagating equation for the electric field $E$, linearly polarized along $x$ and propagating along $z$ according to the wave equation
\begin{equation}
\partial_{tt}E+{{1}\over{\epsilon_{0}}}\partial_{tt}P=c^2 \partial_{zz}E-\sigma \partial_{t}E \, ,
\label{Wave-eq}
\end{equation}
where $P(z,t)$ is the dielectric polarization, $\epsilon_0$ and $c$ are the dielectric constant and the speed of light in vacuum, respectively, and $\sigma$ represents generic losses, which can -- at this stage -- correspond to finite electrical conductivity or to losses through the cavity mirrors.  The field (and thus, the medium's polarization) is assumed to take the form of a plane wave in the transverse direction.  The medium's polarization $P$ is fully described by the susceptibility function $\chi(\omega,N)$ through the Fourier transform 
\begin{equation}
\widehat{P}(\omega)=\epsilon_{0} \chi(\omega,N) \widehat{E}(\omega) \, .
\label{susceptibility}
\end{equation}
Finally, the carrier density obeys
\begin{equation}
\partial_{t} N=\gamma \left(N_{p}-N\right)+D \partial_{zz}N+{{2}\over{\hbar \omega_{c}}}E \partial_{t}P \, ,
\label{population}
\end{equation}
where $\gamma$ is the carrier's relaxation constant, $N_{p}$ the pump parameter, $D$ the diffusion constant, $\hbar$ Planck's constant, and $\omega_{c}$ is the oscillation frequency. 

The form  chosen for the constituent equations, eqs.~(\ref{Wave-eq}--\ref{population}), is entirely general and does not include the usual approximations (RWA, SVEA).  This choice is consistent with the one made in the description of the self-induced ac Stark shift in lasers~\cite{Gil2011} which, at variance with all other laser models, has proven to correctly describe the dependence of the laser oscillation frequency on the field strength. 

The laser description which we adopt is strictly valid near threshold, with the same restrictions which apply to the CGLE for the laser~\cite{Coullet1989}.  Since the dynamics is controlled by the evolution of the physical variables near the bifurcation point, a careful examination of the topological characteristics of the threshold region is necessary; this can be effectively achieved using codimension-2 (CD2) bifurcation theory~\cite{Iooss1991,Iooss1998}.  The model's derivation is extremely complex and lengthy~\cite{Gillong} and a physical description has already been provided~\cite{Gil2014}.  Here, we need to only retain the crucial points of the derivation which consist in:  a. obtaining scaling factors to allow the two variables ($E$ and $N$) to evolve together; b. expanding the variables ($E$, $N$ and $P$) in power of a small parameter $\epsilon$; c. rescaling time and space to separate slow and fast evolution (both in space and time); d. obtaining the solvability conditions which render the assumptions compatible with the equations of the problem, eqs.~(\ref{Wave-eq}-\ref{population}).  The solvability conditions provide the ``slow" differential equations to be satisfied by the new variables:
\begin{eqnarray}
\label{modela}
\partial_{\widetilde{T}}F & = & -V \partial_{\widetilde{Z}} F+\epsilon^{2}c_{0}SF 
\\ \nonumber
& & +\epsilon^{10}c_{1}\partial_{\widetilde{Z}\widetilde{Z}}F+i \epsilon^{12}c_{2i}S \partial_{\widetilde{Z}}F \, ,
\\
\label{modelb}
\partial_{\widetilde{T}}S & = & \widetilde{\mu}-S-4\widetilde{\sigma}\vert F \vert^2+\epsilon^{10}\widetilde{D}\partial_{\widetilde{Z}\widetilde{Z}}S \, ,
\label{ourEquations}
\end{eqnarray}
where we have introduced slow space ($Z$) and time ($T$) coordinates, and $S$ and $F$ represent the leading order amplitudes in the expansion of population and field, respectively (in other words, the order parameters), $\tilde{\mu}$ is the normalized distance from threshold, and the constants $V$, $c_0$, $c_1$ and $c_{2i}$ are defined as follows
\begin{eqnarray}
V = {{2c^2k_{c}^2}\over{\omega_{c}^2 \chi_{\omega}}} \, ,
& \qquad \qquad & 
c_{0} = {{\chi_{iN}}\over{\chi_{\omega}}} \left(1-i \alpha\right) \, , \\
c_{1} = {{-V^2\chi_{i\omega\omega}}\over{2\chi_{\omega}}}\left(1-i \beta \right) \, ,
& \qquad &
c_{2i} = \Im m \left({{-i \chi_{\omega\omega} V c_{0} }\over{\chi_{\omega}}}+{{\chi_{\omega N} V}\over{\chi_{\omega}}}\right) \, , 
\end{eqnarray}
where $k_c$ is the critical eigenvector, $c$ the speed of light in vacuum, $\mathcal{R}e\{c_0\}$$>0$ and $\mathcal{R}e\{c_1\}$$>0$ (by construction), $\alpha$ $=$ $\chi_{rN} / \chi_{iN}$ is the usual alpha-factor and $\beta$ is a new, real function defined by
\begin{equation}
\label{betadef}
\beta={{\chi_{\omega}}\over{V \chi_{i\omega\omega}}}\left({{\chi_{r\omega\omega}V}\over{\chi_{\omega}}}-1\right) \, ,
\end{equation}
with the following definitions for the dimensionless partial derivatives:
\begin{equation}
\chi_{\omega} = \left(\gamma \omega_{c} \right)^{{1}\over{2}} {{\partial \chi}\over{\partial \omega}} \, , \qquad
\chi_{\omega\omega} = \left(\gamma \omega_{c} \right) {{\partial^2 \chi}\over{\partial \omega^2}} \, , \qquad
\chi_{N} = N_{pc} {{\partial \chi}\over{\partial N}} \, ,
\end{equation}
and the subscripts $r$ and $i$ representing real and imaginary parts.

A linear stability analysis for the spatially homogeneous solution ($S = 0$, $F = \sqrt{\tilde{\mu}/4 \tilde{\sigma}}$) gives for the eigenvalue associated with the time-translation symmetry~\cite{Gil2014,Gillong}
\begin{eqnarray}
\label{eigenv}
\Re e \left\{ \lambda_{\phi} \right\} & = & l_2 k^2 + l_4 k^4 + {\cal O}(k^6) \, , \\
l_4 & < & 0 \, , \\
\label{alphabeta}
l_2 & \propto & - (1 + \alpha \beta) \, ,
\end{eqnarray}
where, as in Section~\ref{SMmodel}, we have used an expansion in power of the perturbation wavevector $k$.  When $l_2 > 0$ (i.e., $(1 + \alpha \beta) < 0$) then a phase-unstable regime appears, with possible cyclic oscillations in the laser frequency, as in the experiments~\cite{Yacomotti2004,Furfaro2004,Tanguy2006}.

\begin{figure}
\begin{center}
\begin{tabular}{c}
\includegraphics[height=7cm]{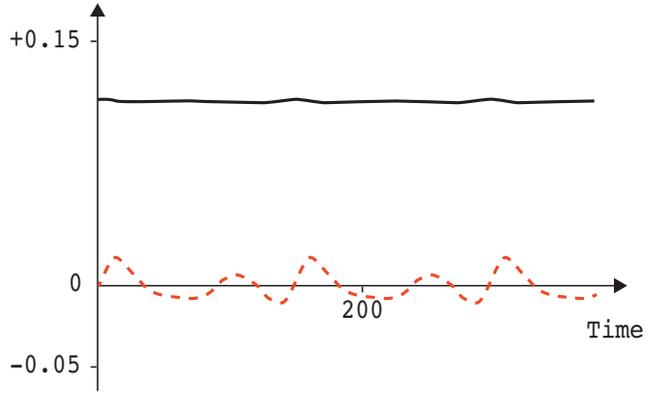}
\end{tabular}
\end{center}
\caption[fig01] 
{\label{fig01}
Numerical simulation of eqs.~(\ref{modela}-\ref{modelb}) in the phase instability regime with $\epsilon$$=$$0.501$, $V$$=$$1$, $c_{0}$$=$$0.05+i 0.01$, $c_{1}$$=$$1.2-i12$, $c_{2i}$$=$$-10$, $\mu$$=$$0.1$, $\sigma$$=$$2$, $D$$=$$1$ and $\chi_{r}$$=$$3$. The space and time increments are $0.05$ and $0.1166$, respectively. The dashed line corresponds to the temporal evolution of the e.m. field frequency and appears to be periodic, though asymmetric. The continuous line represents the time evolution of the intensity $\vert F\vert^2$ and is nearly constant.}
\end{figure}

\section{Predictions of CD2 model}
We have performed numerical simulations of eqs.~(\ref{modela}-\ref{modelb}) using a standard fourth-order Runge-Kutta algorithm for the temporal scheme and a sixth-order finite-difference method to approximate the spatial derivatives. Given the long relaxation time scales expected from the phase dynamics, we have always been careful to converge to the asymptotic regime. 

\begin{figure}
\begin{center}
\begin{tabular}{c}
\includegraphics[height=7cm]{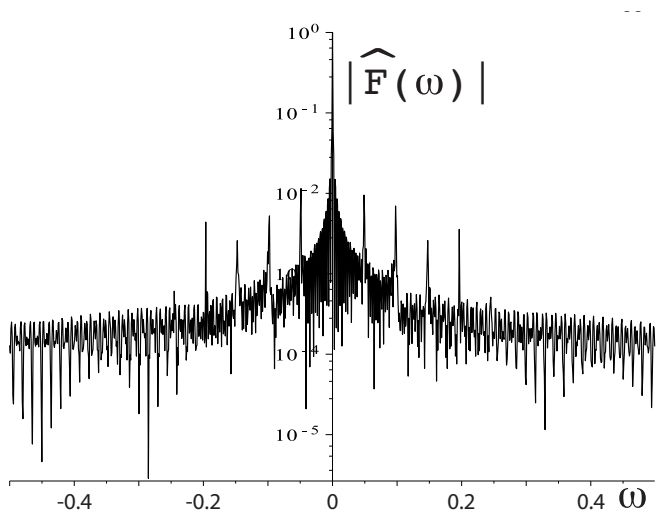}
\end{tabular}
\end{center}
\caption[fig02]
{\label{fig02}
Numerical simulation of eqs.~(\ref{modela}-\ref{modelb}) in the same parameter regime as in Fig.~\ref{fig01}. The figure displays the asymptotic power spectrum of $F$ in log-linear scale.
}
\end{figure}

The deterministic multimode dynamics experimentally observed~\cite{Yacomotti2004,Furfaro2004,Tanguy2006} is then satisfactory reproduced. The total intensity (rather than the {\it incoherent sum}) is nearly constant in time (Fig.~\ref{fig01}, continuous line) while the few cavity modes involved in the dynamics (Fig.~\ref{fig02}) are visited in succession in the temporal evolution (Fig.~\ref{fig01}, dashed line).

\section{Comparison between the two models}
In the exact form in which they are presently expressed, it is not straightforward to compare eqs.~(\ref{SerratEquations1}-\ref{SerratEquations3}) with eqs.~(\ref{modela}-\ref{modelb}) because the former is expected to describe the dynamics for all regimes of parameters while the latter is valid only in the neighborhood of the laser bifurcation threshold. Hence $N$ stands for the true carrier density in eqs.~(\ref{SerratEquations1}-\ref{SerratEquations3}) while $S$ is the deviation from the critical transparency value in eqs.~(\ref{modela}-\ref{modelb}). Nevertheless, some straightforward algebraic manipulations help drawing the comparison.  

Let us first assume $B \equiv 0$ in eqs.~(\ref{SerratEquations1}-\ref{SerratEquations3}) and rewrite $j$$=$$\gamma_{n}N_{p}$. Then the straightforward solution $N$$=$$N_{p}$ and $F$$=$$0$ is linearly stable provided that $N_{p}$ $<$ $N_{pc}$ $=$ $1+\gamma_{n}$. Substituting 
\begin{equation}
N_{p}=N_{pc}+\mu \, ,
\qquad \qquad
N=N_{pc}+S \, ,
\qquad \qquad 
F={\cal F} e^{-i {\cal K} \alpha N_{pc} t}
\end{equation}
in eqs.~(\ref{SerratEquations1}-\ref{SerratEquations3}) leads to
\begin{eqnarray}
\label{SMrewritten1}
\partial_{t}{\cal F} & = & -\partial_{z}{\cal F}+{\mathcal G}\left( S \left( 1-i\alpha \right)+\gamma_{int}G_{d}\partial_{zz}\right){\cal F} \\
\nonumber
 & &+{\mathcal G} S \partial_{zz}{\cal F} \, , \\
\label{SMrewritten2}
\partial_{t}S & = & \gamma_{n}\left(\mu-S\right)-\gamma_{i}\vert {\cal F}\vert^2+d \partial_{zz}S \\
\nonumber
& &-\left(\mu+S\right) \vert {\cal F}  \vert^2 \, ,
\label{SMrewritten}
\end{eqnarray}
where we have grouped on the first line of each of the previous equations the terms which are common to both  eqs.~(\ref{SerratEquations1}-\ref{SerratEquations3}) and eq.~(\ref{modela}-\ref{modelb}). $\mu$ represents the distance from threshold.   Comparing the two descriptions we remark that: \begin{enumerate}
\item The terms ${\mathcal G} S \partial_{zz}{\cal F}$ and $\left(\mu+S\right)  \vert {\cal F}  \vert^2$, which are present in eqs.~(\ref{SMrewritten1}-\ref{SMrewritten2}) and not in eqs.~(\ref{modela}-\ref{modelb}), do appear in the course of the derivation of eqs.~(\ref{modela}-\ref{modelb}) but have been neglected because of their smallness in the neighborhood of the laser threshold \cite{Gillong}.
\item The term $i c_{2i} S \partial_{z} F $ is present in eq.~(\ref{modela}) but not in eq.~(\ref{SMrewritten1}).  This term describes how the selected wave vector is shifted with the carrier density. Note that this term does not enter into the Benjamin-Feir phase instability criterium.
\item The coupling between phase and amplitude through the $\alpha$ factor is the same for both description.
\item In eqs.~(\ref{modela}-\ref{modelb}), the coefficient in front of $\partial_{zz}F$ is complex, while it is purely real in eq.~(\ref{SerratEquations1}-\ref{SerratEquations3}).  In other terms, $\beta$ is strictly zero (by construction) in the SM model, which  definitively excludes the possibility of appearence of a phase instability regime.
\end{enumerate}

We conclude that, although the two models (SM and CD2) are structurally identical -- i.e., contain the same leading terms --, the nature of one of their parameters is different.  While the coefficient in front of the diffraction term (second line of eq.~(\ref{SMrewritten1})) is complex in the CD2 model, in the SM model it is strictly real.  This is exactly the reason why a phase instability cannot occur in the SM model, but can appear in the CD2 one.  

As a general remark, one should notice that for a coefficient (or one of its complex parts) to be exactly zero, an underlying profound physical selection process must exist (e.g., a symmetry of the problem or the existence of a conservation law).  In general, coefficients should possess both components:  their values will determine whether the corresponding process is sizeable or neglibible (here the phase instability, elsewhere, e.g., the dependence of the laser frequency on the electric field strength~\cite{Gil2011}).

Up until here we have discussed the analytical comparison between the two models using the restriction, for the SM one, to a single direction of propagation.  This assumption bears no weight on our conclusions, since the two counterpropagating waves are coupled through amplitude and not phase terms.  Thus, considering the full bidirectional nature of the SM model does not change the results of this analysis.  This is confirmed by the analysis of Section~\ref{SMmodel} which proves that the stability properties are generic, independently of the value of $\theta$ (eqs.~(\ref{coeff12}-\ref{coeff3})), thus of the relative amplitude of the two counterpropagating e.m. field components.

\section{Conclusions}

The description of the semiconductor laser based on the interaction between the full e.m. field and the lasing medium, introduced by Serrat and Masoller~\cite{Serrat2006}, has represented a innovative step beyond the modal decomposition, traditionally used in the description of the longitudinal multimode operation of semiconductor lasers.  However, the approximations introduced to obtain a manageable model~\cite{Serrat2006} remove all possibilities for observing a true phase instability.  The very slow convergence of the computer codes, owing to the presence of very long time scales, led to the false conclusion that long-term phase dynamics could be reproduced in this way~\cite{Serrat2006}.

The full mathematical description that our model provides, based on a complete Codimension 2 expansion, offers the advantage of a proper and global handling of the various physical processes (FWM, CS, SS) in the neighbourhood of threshold, without ressorting to {\it ad hoc} choices to introduce a physical coupling mechanism.  

The {\it ab initio} derivation of the equations, avoiding all the usual approximations -- as already done~\cite{Gil2011} --, has allowed us to discover the existence of a phase-instability which arises through a {\it new fundamental parameter}, named $\beta$-parameter, which in conjunction with the traditional $\alpha$-parameter of semiconductor lasers,  controls the growth of the phase-eigenvalue and maps the semiconductor laser problem into the Benjamin-Feir instability~\cite{Benjamin1967} of fluid dynamics:  a problem whose parameter space has been carefully explored and characterized~\cite{Manneville1985,Chate1994}.  The existence of the $\beta$-parameter and its joint-control, with $\alpha$, of the instability is the probable reason why semiconductor lasers as different as MQWs and QDLs may present similar dynamics.  This point is left for further investigation. 

We have shown that the phase instability operates in a way which is similar to the random excitation due to noise, but in an entirely deterministic kind of description.  This comes from the width of the unstable interval of the Lyapunov exponents in the spectrum of the eigenvalues~\cite{Manneville1985}.  Thus, our description reconciles the phenomenological deterministic description of the problem~\cite{Yacomotti2004,Furfaro2004} with the stochastic one which couples the mode through the projection of noise on the mode ensemble~\cite{Ahmed2002}.  At the same time, however, our results open new questions on the relative role of the phase instability (as {\it noise source}) compared to the physical, unavoidable, sources of noise~\cite{Osinski1987}.  Answering these questions requires further work; at the present stage we can only point out their existence.

\end{document}